\documentclass[twocolumn,aps,epsfig,showpacs,floatfix,superscriptaddress]{revtex4}
\usepackage{epsfig}
\usepackage{graphicx}
\usepackage{dcolumn}
\begin{document}


\title{Influence of nuclear structure on sub-barrier hindrance in Ni+Ni fusion}
\author{C.L. Jiang}
\author{ K.E. Rehm}  
\author{R.V.F. Janssens}  
\author{H. Esbensen}  
\author{I. Ahmad}  
\author{B.B.~Back}\affiliation{Physics Division, Argonne National Laboratory, Argonne, IL 60439}  
\author{P. Collon}\affiliation{University of Notre Dame, Notre Dame, IN 46556}
\author{C.N. Davids}\affiliation{Physics Division, Argonne National Laboratory, Argonne, IL 60439}  
\author{J.P. Greene}
\author{D.J. Henderson}
\author{G. Mukherjee} \altaffiliation{Present address: Saha Institute of Nuclear Physics, Kolkata, India}
\author{R.C.~Pardo}\affiliation{Physics Division, Argonne National Laboratory, Argonne, IL 60439}  
\author{M. Paul}\affiliation{Hebrew University, Jerusalem 90914, Israel}
\author{T.O. Pennington} \altaffiliation{Deceased} 
\author{D. Seweryniak} 
\author{S. Sinha}  
\author{Z. Zhou} \affiliation{Physics Division, Argonne National Laboratory, Argonne, IL 60439}

\date{\today}

\begin{abstract}
Fusion-evaporation cross sections for $^{64}$Ni+$^{64}$Ni have been measured down to the 10 nb level. For fusion between two open-shell nuclei, this is the first observation of a maximum in the $S$-factor, which signals a strong sub-barrier hindrance. A comparison with the $^{58}$Ni+$^{58}$Ni, $^{58}$Ni+$^{60}$Ni, and $^{58}$Ni+$^{64}$Ni systems indicates a strong dependence of the energy where the hindrance occurs on the stiffness of the interacting nuclei. 
\end{abstract}
\maketitle

In the simplest picture, fusion reactions between two heavy ions at low
energies are governed by penetration through the interaction barrier followed by absorption inside the barrier. With the discovery of sub-barrier fusion enhancement, it was realized that coupling between fusion and other degrees of freedom creates a multi-dimensional potential barrier resulting in increased fusion probabilities \cite{summ}. A better understanding of the fusion process followed the introduction of ``experimental barrier distributions'' \cite {rowl}. Many detailed distributions of fusion barriers, originating from experimental measurements and from coupled-channels calculations, have been obtained in recent years. However, the study of the interplay between the interaction barrier, the tunneling process and the absorption has been restricted mostly to energies in the vicinity of the barrier because of difficulties with measuring fusion cross sections below the $\sim$100 $\mu$b level. In this energy region coupled-channels calculations do reproduce 
the experimental data quite well. 

Measurements of fusion reactions between heavy ions at extreme sub-barrier energies are also of interest for reaction mechanism studies in astrophysics and for the synthesis and study of super-heavy elements. It has recently been pointed out \cite {jiang0} that, at extreme sub-barrier energies, the fusion cross sections exhibit a behavior that is 
different from the predictions of coupled-channels calculations. A much 
steeper falloff was observed at energies far below the Coulomb barrier for the 
systems 
$^{58}$Ni + $^{58}$Ni \cite {beck1},
$^{90}$Zr + $^{90}$Zr, $^{92}$Zr and $^{89}$Y \cite {kell}, 
and $^{60}$Ni + $^{89}$Y \cite {jiang0}.
This feature is emphasized in the so-called logarithmic derivative,
$L(E)=d(\ln\sigma E)/dE$, which continues to increase
with decreasing energies \cite{jiang0}. 

Recently, a detailed coupled-channels analysis has been performed  for the system $^{60}$Ni + $^{89}$Y \cite {jiang1}, which was measured down to the 100 nb level, in order to confirm the conclusions in Ref. \cite {jiang0}. Another representation, in terms of the so-called {\it S}-factor, has also been introduced by these authors \cite{jiang1} to study this behavior further. In the {\it S}-factor representation, the steep falloff in cross sections translates into a maximum of the {\it S}-factor. The energies where the maximum occurs for these five systems
can be parameterized with a simple empirical formula. The parameterization, which was derived for rather stiff nuclei, also provides an upper limit for reactions involving softer nuclei. 

It is of great interest to investigate whether the behavior observed in Refs. \cite {jiang0,jiang1} persists also in fusion between open-shell nuclei. We have, therefore, studied fusion-evaporation reactions in the system $^{64}$Ni + $^{64}$Ni. Previous measurements \cite{beck2,acke} obtained data down to about 20 $\mu$b. In this energy region, the {\it S}-factor of the system $^{64}$Ni + $^{64}$Ni has not quite reached a well-defined maximum, but starts to deviate at the lowest energies from the calculations based on the coupled-channels formalism \cite {jiang1}. 

From an experimental point of view, $^{64}$Ni + $^{64}$Ni is a good choice for precise measurements of fusion cross sections because there are no complications arising from reactions on small amounts of heavier isotopic contaminants in the target, or from lower-Z isobaric components in the beam, which can dominate the low-energy yields at extreme sub-barrier energies \cite{jiang0}.

The experiment was performed at the Argonne super-conducting linear accelerator ATLAS using the Fragment Mass Analyzer (FMA) \cite {dav1} to identify and measure fusion evaporation products. Metallic $^{64}$Ni of 98.02\% isotopic purity and an areal thickness of 82 $\mu$g/cm$^2$ was evaporated onto a thin carbon foil (100 $\mu$g/cm$^2$, facing the FMA). The 2\% impurities in the target were all lighter Ni isotopes, which contribute only insignificantly to the measured cross section because of a higher interaction barriers. The abundances were: $^{58}$Ni, 0.97\%, $^{60}$Ni, 0.57\%, $^{61}$Ni, 0.05\%, and $^{62}$Ni, 0.39\%. A carbon foil (10 $\mu$g/cm$^2$) was placed 2 cm downstream from the target to reset the charge state distribution of the recoiling residues. The maximumbeam current was about 100 pnA. The detection technique for the residues and the data analysis used in the experiment have been described previously \cite {jiang0}.

The measurement was performed in the energy range of 171 to 220 MeV. Residues with two charge states were detected simultaneously at the focal plane of the FMA. Complete charge state distributions of evaporation residues were measured at three energies, 217, 190 and 180 MeV. 
These data, together with a parameterization \cite{shima}, are sufficient to determine with the desired accuracy the charge state fractions of the detected evaporation residues at each beam energy. 
Compared to our earlier experiment \cite{jiang0}, the anode in the first electric dipole of the FMA was replaced by a split anode \cite {dav3}. As a result, the background induced by scattered beam particles was greatly reduced, allowing measurements of fusion-evaporation cross sections down to a few tens of nanobarns.
\begin{figure}
\epsfig{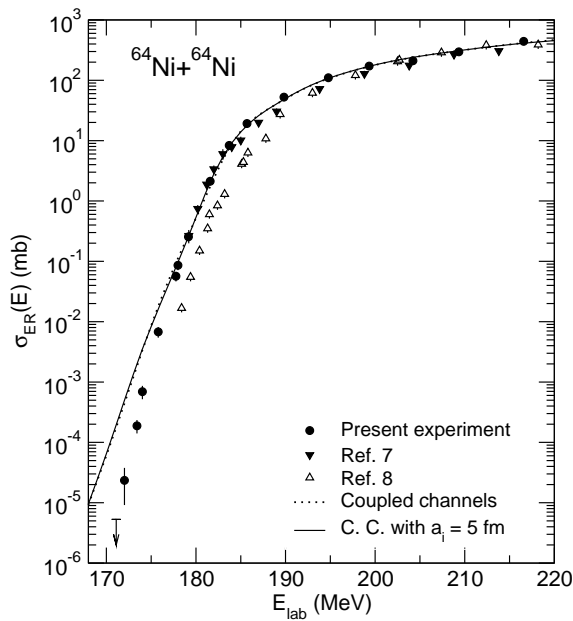}
\caption{Experimental evaporation residue cross sections $\sigma (E)$ 
(solid circles) plotted as a function of laboratory energy $E_{lab}$ for the system $^{64}$Ni + $^{64}$Ni. The incident energies have been corrected for the finite target thickness. The uncertainties in the cross sections for many points are smaller than the size of the symbol. Previous measurements from Refs. {\cite {beck2,acke}} are shown as solid  and open triangles, respectively. The dashed and solid curves represent coupled-channels calculations, which are fit to the high energy part of the present data (see text for details).}
\end{figure}

The present results are shown in Fig. 1, together with the cross sections obtained in Refs. \cite {beck2,acke}. The incident energies have been corrected for the finite target thickness (including the effect of sharp changes in the excitation function with energy). At the lowest center-of-mass energy, $E$=85.55 MeV, no evaporation residues were observed in an eight hour run. The upper limit in Fig. 1 represents the cross section corresponding to one count. Our measurements are in good agreement with the results of Ref. ~\cite{beck2}, but are shifted by about $\Delta E_{\it lab}\sim$1.5 MeV towards lower energies compared to the results of Ref. \cite{acke}. The solid and dotted curves are the results of coupled-channels calculations described below.

Under the assumption that the discrepancy between the present results and those of Ref. \cite {acke} is caused by a shift in beam energy by one of the experiments, we have studied further the beam energy calibration of the ATLAS accelerator. Thus, we have performed beam energy measurements using several heavy ion species in the energy range of 1 - 10 MeV/u. The determination of the beam energy at ATLAS is accomplished with a resonant detection time-of-flight system \cite {rich}. Such a system can be referenced to ``absolute'' parameters such as the physical distance between the resonant detectors and the electronic delay of signals from various components in the system. A more precise calibration may be obtained by comparing the time-of-flight with a direct measurement of the magnetic rigidity of the beam by passing it through an Enge magnetic spectrograph, for which the focal plane detector was in turn calibrated by means of $^{228}$Th, $^{238}$Pu, $^{241}$Am, $^{244}$Cm, and $^{249,250}$Cf sources \cite {rytz}. Beams of $^{16}$O, $^{78}$Kr, $^{60}$Ni and $^{64}$Ni were used with the magnetic spectrograph placed at zero degrees. 
This calibration of the ATLAS energy measurement system showed that the beam energies quoted in the present work have an accuracy of about 0.1\% ($\sim 200$ keV), {\it i.e.} significantly lower than the discrepancy between the two cross section measurements in question. 
It should be noted that the systematic energy differences are not
important for the main physics point of this Letter.

\begin{table}
\caption{Energies and transition probabilities for $^{64}$Ni states included in the coupled-channels calculations.}
\vspace{1mm}
\begin{tabular} {ccccc}
\hline
\hline
$\lambda^\pi$ &  E$_x$  & B(E$\lambda$)       &  $\beta_\lambda^{Coul}$ &$\beta_\lambda^{Nucl}$\\
              &  (MeV)  & ($e^2b^{2\lambda}$) &\\
\tableline
   $2^+$      &  1.346  &   650\footnote{Ref. ~\cite{71ChZT}}  & 0.165&                  0.185\footnote{Ref. ~\cite{Flem}} \\
   $3^-$      &  3.560  & 20400\footnote{Ref. \cite{Braun}}           & 0.193 & 0.200 \\
\hline
\end{tabular}
\end{table}

Coupled-channels calculations have been performed previously \cite {henn3} for $^{64}$Ni+$^{64}$Ni  by fitting the data of Ref. ~\cite {beck2}. The present calculations are fitted to the new data with the nuclear structure input given in Table I. The full calculations include $2^+$ and $3^-$ one-phonon excitations, the mutual excitation, and the two-phonon quadrupole excitation estimated within a vibrational model. The ion-ion interaction parameters are as follows: potential $V_0$ = 75.98 MeV, diffuseness $\it a$ = 0.676 fm and radius $R$ = 9.52 fm. Here a radius shift $\Delta R$ = 0.10 fm (with reference to the systematic radius \cite {brog}) has been adjusted to minimize the $\chi^2$ deviation from the data at high energies. The results are given in Fig. 1 as a dotted curve in comparison with the experimental data.  A modified coupled-channels calculation, which increases the diffuseness parameter inside the barrier to a value $\it a_i$ and keeps the diffuseness parameter outside the barrier at the original value, was introduced in Ref. \cite {jiang1}. The calculation with $\it a_i$= 5 fm is the solid curve in Fig. 1. It is evident that the experimental cross sections exhibit a steeper falloff below $E_{\it lab}$= 176 MeV than can be accounted for by the coupled-channels calculations.

\begin{figure}
\epsfig{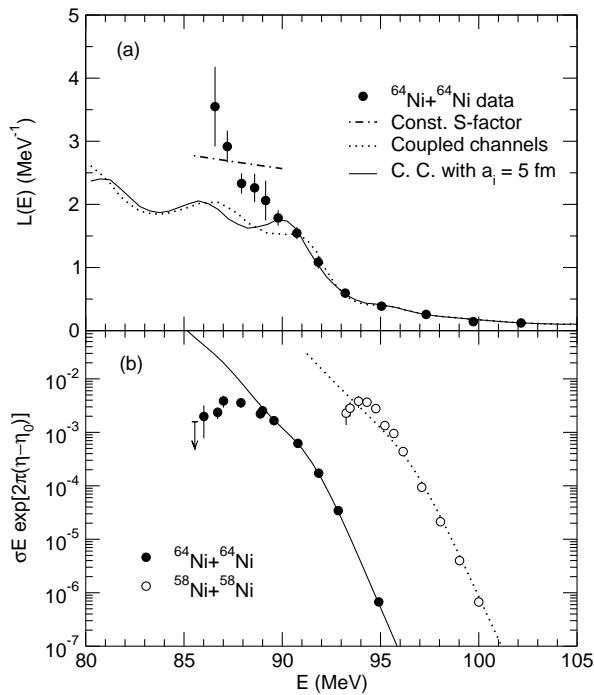}
\caption{
(a) The logarithmic derivative $L(E) = d(\ln\sigma E)/dE$
plotted as function of center-of-mass energy $E$. The solid circles were derived from the data by least-squares fits to three consecutive data points. Coupled-channels calculations are shown as the solid and dotted curves, whereas the dash-dotted curve corresponds to a constant $S$-factor. (b) The $S$-factor for $^{64}$Ni+$^{64}$Ni (solid points) is compared to that for $^{58}$Ni+$^{58}$Ni \cite{beck1} (open points). The solid curve represents a coupled-channels calculation using $a_i$=5 fm for the $^{64}$Ni+$^{64}$Ni system. The dotted curve is a coupled-channels calculation taken from Ref. ~\cite{henn3}. The parameter $\eta_0$ used to bring different fusion systems onto the same scale are $\eta_0$=75.23 and $\eta_0$=69.99 for the $^{64}$Ni+$^{64}$Ni and $^{58}$Ni+$^{58}$Ni systems, respectively.}
\end{figure}

The experimental logarithmic derivatives, presented as solid points in Fig. 2a, exhibit an increase with decreasing energy while the coupled-channels calculations are found to increase only modestly. The dash-dotted curve in Fig. 2a represents an $s$-wave transmission for a pure point charge Coulomb potential or a constant $S$-factor. As shown in Ref. \cite {jiang1}, the logarithmic derivative is in this case given by
\begin{equation}
 L_{\it CS}(E)=\frac{\pi\eta}{E},
\end{equation}
where $\eta$ is the Sommerfeld parameter. At the lowest energies, all calculated curves are nearly parallel, and they are unable to describe the general behavior of the experimental data. This observation indicates that a substantial component, yet to be identified, is missing in the description of the reaction.

 The $S$-factor representation of the $^{64}$Ni+$^{64}$Ni data (solid points) is shown together with the coupled-channels calculation (using $\it a_i$= 5 fm, solid curve) in Fig. 2b. A clear maximum of the $S$-factor is observed in the experimental data, but not in the calculation.  In Ref. \cite {jiang1}, an extrapolated value for the location of the maximum of the $S$-factor, $E_s$ = 89.0 MeV, was obtained using the data of Ref. \cite{acke}. Based on the present experiment, the measured value is 87.7 MeV, with the difference mainly due to the systematic shift of the two excitation functions noted above. The location in energy of this maximum occurs at the crossing point of experimental logarithmic derivatives and the $L_{\it CS}(E)$ curve. For comparison we show also the $S$-factor for the $^{58}$Ni+$^{58}$Ni system of Ref. \cite{beck1}. We note that the $S$-factor maximum in the latter case occurs at a significantly higher energy, $E_s$=94 MeV. The location of these $S$-factor maxima  will be discussed further below.
\begin{figure}
\epsfig{file=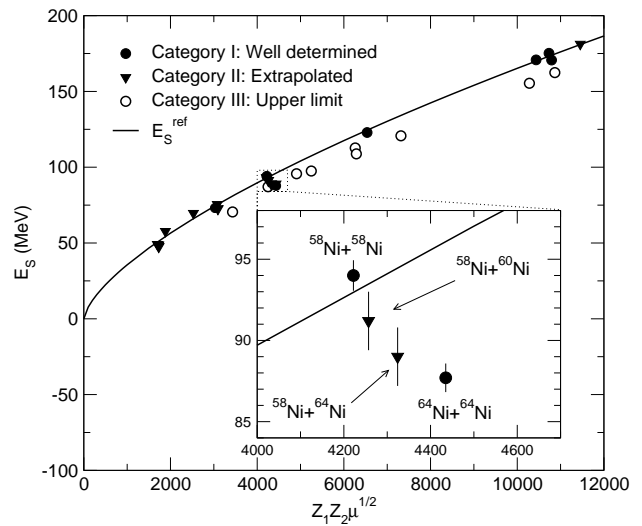,width=\columnwidth}
\caption{Systematics of the energy $E_s$, where the {\it S}-factor
has a maximum as a function of the parameter, $Z_1Z_2\sqrt{\mu}$. The solid curve represents the empirical expression given by Eq. 1. Solid circles were obtained for systems that exhibit a clear maximum in the {\it S}-factor, {\it i.e.} so-called category I systems \cite{jiang1}. Solid triangles were obtained by extrapolating the logarithmic derivative to the value for a constant {\it S}-factor (category II systems). Open circles show the lowest measured energy, $E_{min}$, for those systems where no sign of a maximum in the {\it S}-factor has been found so far (category III systems). The insert gives an expanded view of the results for Ni+Ni reactions.}
\end{figure}
\begin{table}
\caption{The parameter $Z_1Z_2\sqrt{\mu}$ and the location of the maximum of the $S$-factor are listed for different systems.}
\vspace{1mm}
\begin{tabular} {ccccc}
\hline
\hline
Category &     System        & $Z_1Z_2\sqrt{\mu}$  & $E_S$ (MeV)& Ref.\\
\tableline
I        & $^{32}$S+$^{89}$Y   & 3027         &   73.1           &\cite{mukh}\\
II       & $^{34}$S+$^{89}$Y   & 3095         &   72.6           &\cite{mukh}\\
II       & $^{28}$Si+$^{58}$Ni & 1704         &   49             &\cite{stef2}\\
II       & $^{28}$Si+$^{62}$Ni & 1722         &   48.6           &\cite{stef2}\\
II       & $^{28}$Si+$^{64}$Ni & 1730         &   47.6           &\cite{stef2}\\
I        & $^{58}$Ni+$^{58}$Ni & 4222         &   94             &\cite{beck1}\\
II       & $^{58}$Ni+$^{60}$Ni & 4257         &   92             &\cite{stef3}\\
II       & $^{58}$Ni+$^{64}$Ni & 4324         &   89             &\cite{beck2}\\
I        & $^{64}$Ni+$^{64}$Ni & 4435         &   87.7           &Present\\
\hline
\end{tabular}
\end{table}
It has recently been pointed out \cite{jiang1} that the center-of-mass energy, $E_s^{\it ref}$, of the $S$-factor maximum observed in five fusion systems involving ``stiff'' nuclei (category I systems, table I in Ref. \cite{jiang1}) is well approximated by
\begin{equation}
\label{empri}
E_s^{\it ref} = 0.356\left(Z_1Z_2\sqrt{\mu}\right)^{\frac{2}{3}} \rm{(MeV)},
\end{equation}
where $\mu=A_1A_2/(A_1+A_2)$ is the reduced mass of the system. This expression corresponds to a value of $L_{CS}$ =2.33 MeV$^{-1}$. The systems $^{32,34}$S+$^{89}$Y \cite{mukh}, $^{28}$Si+$^{58,62,64}$Ni \cite{stef2},$^{58}$Ni+$^{60}$Ni \cite{stef3}, and $^{58}$Ni+$^{64}$Ni \cite{beck2} in addition to the present measurement of $^{64}$Ni+$^{64}$Ni have been added to the systematics. The results from this analysis are listed in Table II. We find that the $^{32}$S+$^{89}$Y system exhibits an $S$-factor maximum at $E_{S}$=73.1 MeV and it is, therefore, labeled as a category I system, see Ref. \cite{jiang1}. This system falls on the $E_s^{\it ref}$ curve, which is consistent with the expectation for systems involving rather ``stiff'' nuclei in the entrance channel. 

It is interesting to compare the four systems
$^{58}$Ni+$^{58}$Ni \cite{beck1},
$^{58}$Ni+$^{60}$Ni \cite{stef3},
$^{58}$Ni+$^{64}$Ni \cite{beck2}, and
$^{64}$Ni+$^{64}$Ni in order to study the progression from ``stiff'' to ``open-shell'' nuclei in the entrance channel. The energies $E_s$ plotted vs. $Z_1Z_2\sqrt{\mu}$ for these four systems are shown as an insert in Fig. 3. The solid circles are for $^{58}$Ni+$^{58}$Ni and $^{64}$Ni+$^{64}$Ni for which $E_s$ values were obtained with an uncertainty of $\sim 1 \%$. The triangles represent the systems $^{58}$Ni+$^{60}$Ni and $^{58}$Ni+$^{64}$Ni for which $E_s$ were obtained by extrapolations of the logarithmic derivative to the crossing point of $L_{CS}(E_S)$. The accuracy of this procedure is estimated to be $\sim$2\%. It is evident that the deviation of $E_s$ from $E_s^{\it ref}$ is related to the neutron number of the colliding nuclei in the entrance channel, which also reflects the stiffness of the systems. For the ``soft'' $^{64}$Ni+$^{64}$Ni system the measured value of $E_s$=87.7 MeV is about 9\% lower than the value of $E_s^{\it ref}$=96.1 MeV, which is expected based on the systematics for ``stiff'' nuclei. Note that the interaction barrier is reduced by only 3\%. A similar, but less accurately determined, behavior is observed for the systems $^{40}$Ar + $^{144,148,154}$Sm \cite{reis}, and  $^{90,92,96}$Zr \cite{kell}.

In conclusion, we have  measured the fusion excitation function for $^{64}$Ni+$^{64}$Ni down to a cross section level of 10 nb and have observed a strong fusion hindrance at extreme sub-barrier energies. In comparison with data for $^{58}$Ni+$^{58}$Ni, we find that the onset of the sub-barrier fusion hindrance in $^{64}$Ni+$^{64}$Ni occurs 8.4 MeV lower in center-of-mass energy, whereas a 2 MeV higher energy was expected on the basis of systematics. This effect appears to be associated with the nuclear structure of the interacting nuclei, $^{64}$Ni being ``softer'' than $^{58}$Ni. At this point there is a clear experimental observation of sub-barrier suppression and of its dependence on the structure of the interacting nuclei. No satisfactory theoretical explanation of this effect has been proposed thus far. By measuring the fusion process to ever lower sub-barrier energies the dependence on the interaction potential at shorter ion-ion distances is being probed in a way that may reveal inadequacies of the present assumptions ~\cite{Dasso}. Further work, both experimental and theoretical, is required to reach an understanding of this phenomenon.

This work was supported by the U.S. Department of Energy, Nuclear Physics
Division
under Contract No. W-31-109-ENG.

\end{document}